# Adventures with Micro- and Nanofabricated Devices for Chemical and Biochemical Measurements; Early Days to Present!

*J. Michael Ramsey - University of North Carolina*

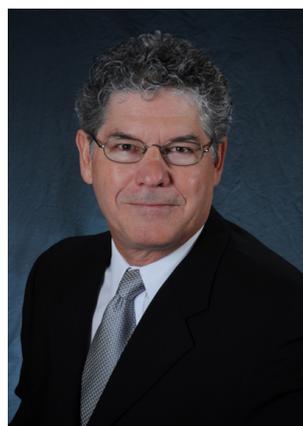

## Biography

*Dr. J. Michael Ramsey holds the Minnie N. Goldby Distinguished Professor of Chemistry Chair at the University of North Carolina - Chapel Hill. In addition, he is a member of the faculty in the Departments of Biomedical Engineering and Applied Physical Sciences, and the Carolina Center for Genome Sciences. His present research interests include microfabricated chemical instrumentation, micro- and nanofluidics, single molecule DNA characterization, single cell assays, point-of-care clinical diagnostic devices, and highly miniaturized mass spectrometry. He is a member of the National Academy of Engineering and a Fellow of the Optical Society of America, the American Chemical Society, and the American Institute for Medical and Biological Engineering. A partial list of awards includes a senior Alexander von Humboldt Award, the A. J. P. Martin Gold Medal for Separation Science, the Marcel J.E. Golay Award in Capillary Chromatography, the Jacob Heskel Gabbay Award in Biotechnology and Medicine, the American Chemical Society Division of Analytical Chemistry Award in Chemical Instrumentation, the Pittsburgh Analytical Chemistry Award, the American Chemical Society Award in Chromatography, the CASSS Award for Outstanding Achievement in Separation Science, the Ralph N. Adams Award in Bioanalytical Chemistry and three R&D 100 technology awards. Dr. Ramsey is the sole scientific founder of Caliper Technologies (NASDAQ:CALP), renamed Caliper Life Sciences and acquired by PerkinElmer in 2011. He is also the scientific founder of the venture-backed companies 908 Devices Inc., a company developing revolutionary compact mass spectrometry and chemical separations-based products, and Genturi Inc., a genomics tools provider. Prof. Ramsey has published over 300 peer-reviewed papers (H-index = 61) and presented over 500 invited, plenary, or named lectures. In addition, he has over 100 issued and over 40 pending patents.*

## Proceedings

My career in science, as of the late 1980s, had been focused on laser-based chemical measurement strategies and instrumentation. Shortly after starting my career at Oak Ridge National Laboratory (ORNL) in 1979, I read a news brief about work performed at Stanford where a gas chromatograph had been implemented on a 50-mm silicon wafer using micromachining techniques[1]. The Stanford μGC efforts were soon commercialized by Microsensor Technology, Inc. (MTI) but primarily sold products that used conventional open-tubular silica capillary columns rather than silicon micromachined columns due to performance advantages. The etched silicon columns had degraded performance compared to conventional drawn circular cross section capillaries because of mass transport issues in both the gas and stationary phases that were related to the nominally rectangular etched silicon column cross sections. I also followed the work of Prof. James Jorgenson on capillary electrophoresis (CE)[2] as we were contemporaries in graduate school.

The genesis of my notion of implementing CE on a microfabricated substrate was combining the work of Jorgenson and Terry, et al, in the late 1980s; realizing that the mass transport issues associated with the micromachined gas chromatography column do not arise in a CE experiment because of the planar axial velocity vector field generated by electrokinetic transport. I found this notion very intriguing and decided to pursue the experimental realization of such devices. My first proposals on this topic were in 1990 to the Measurement and Controls Engineering Center at the University of Tennessee, a US National Science Foundation supported center, and to the US Department of Energy (DOE) Office of Basic Energy Sciences; the latter agency was supporting my laser research at ORNL. Both proposals were rejected, presumably in part because I had no experience in either micromachining or chemical separations and a lack of a fully developed motivation for such technology. I then visited companies such as Varian, Hewlett Packard, Waters, and Applied Biosystems to promote the idea of micromachined separations technology in 1990-1991 without receiving any financial or moral support. My host, for one of these company visits, suggested that I should focus on laser spectroscopy and forget about microfabricated CE! Finally perseverance was rewarded with an ORNL seed grant in 1991 and a multiyear, multi-person grant from DOE later that same year.

Our earliest work was entirely focused on efficient implementation of CE in glass microfabricated devices. Our first paper focused on strategies for reproducibly injecting short axial extent plugs, the "pinched" injection, of sample into channels and the impact of serpentine channel layouts to minimize device footprint[3] and a sequential paper on the use of pinched injections to realize sub-second CE separations[4]. Soon we realized that even faster separations could be performed and demonstrated sub-millisecond electrophoretic separations[5]. A number of electrokinetic separation techniques, such as electrochromatography and sample stacking, were demonstrated on monolithically fabricated chips using different types of sample materials such as peptides, proteins and nucleic acids. The general advantages of placing electrokinetic separations on these microfabricated devices were that separations could be completed ≈100x faster than conventional capillary-based experiments with no loss in separative performance. Sample volumes consumed in an experiment were also ≈100x smaller but volumes of sample that were required to be loaded onto a device where generally in the range of 1 µL, an interface issue that we coined the "world-to-chip" interface problem that in large part still exists today.

Further advantages of microfabricated implementations of CE are the integration of multiple processes on a device. The first monolithic lab-on-a-chip experiments involved demonstration of CE peptide separations with integrated pre-column[6] and post-column[7] reactors where we also described the "gated" injection technique. A later integrated device involved the digestion of DNA with a restriction endonuclease followed by sizing of the fragments. This device utilized 30 amoles of DNA and took 5 minutes for a complete analysis[8]; this is one of my favorite experiments as we reduced the amount of material needed for such an experiment by more than $10^4$ and the time required by $10^2$.

After early demonstrations of electrokinetic separations, we focused on fabrication[9] and architecture strategies[10] to optimize performance and function. The design and modeling efforts led us to a compact microchip CE design that allowed the generation of over one-million theoretical plates in less than a minute[11] and the realization of a 2D chromatographic – electrophoretic separation able to generate a peak capacity of 5000 in less than 15 min[12]. We also realized that these devices could be used for many applications beyond chemical separations such as enzyme kinetic studies[13], flow cytometry[14], DNA hybridization assays, and single cell kinase assays[15] to name a few.

This early work stimulated numerous patent applications and the issuing of over 80 U.S. and foreign microfluidics related patents to our research group. Many of these patents were foundational for the formation of Caliper Microanalytical Systems, Inc. by Michael Knapp and myself in 1994. A year later we joined forces with the late Lawrence Bock and the three of us formed Caliper Technologies, to which all of these patents were transferred. All of Caliper's microfluidic products utilized our electrokinetic fluid manipulation patents.

Our successes in microfluidics motivated us to consider other forms of chemical measurement instrumentation that could benefit from miniaturization. In 1996 we began a project to investigate the miniaturization of mass spectrometry. We chose ion trap mass analyzers because the physics of these devices scaled most attractively compared to all other types known to us. For example, the mass-to-charge ratio (m/z) resolving power of ion traps does not depend on device dimensions and the charge capacity, which relates to sensitivity and dynamic range, scales with their linear dimension rather than volume. Our early work focused on the use of sub-millimeter scale cylindrical ion traps, 20x smaller than commercial ion traps, that were conventionally machined and operated at buffer gas pressures of ≈1 mTorr[16], or typical ion trap pressures. It was the realization that these miniature ion traps could be operated at elevated pressures that made this technology really interesting[17]! We were eventually able to demonstrate mass spectrometry at pressures exceeding 1 Torr using micromachined ion traps, or a pressure 1000x higher than the highest operating pressure in commercial mass spectrometers. We have demonstrated this "high pressure" operation using He[18] and air[19] buffer gas; the latter being important for the realization of a compact mass spectrometry system. We coined the name High Pressure Mass Spectrometry (HPMS) to describe this mode of performing mass spectrometry. This work has led to twelve issued patents, to date, and was the basis for the formation of another company, 908 Devices Inc. 908 Devices launched the world's first handheld mass spectrometer in 2014 based upon these patents with a weight of 2 kg, including a battery to operate for greater than 5 hours.

We have been investigating transport in nanofluidic scale devices since 2000 with early work focusing on electroosmotic transport[20] and electrokinetic polyelectrolyte transport[21]. We began working on single molecule electronic sequencing in 2004 where we were attempting to linearize ssDNA through nanoconfinement with interrogation of single nucleotides via electron tunneling currents[22]; a very challenging problem that requires critical dimensions of ≈1-2 nm. These efforts led to developments in state-of-the-art focused ion beam milled features[23] with dimensions below 5 nm, and sub-nanometer scale electrodes[24]. The integration of these technologies into fluidic circuits proved quite challenging, thus we relaxed our goals to utilizing nanochannels for performing single molecule DNA physical mapping experiments. Rapid, high-resolution mapping of genomic DNA is still an unmet need in an era where next generation sequencing (NGS) can provide exquisite single nucleotide variation. Tools to measure long-range genomic variation are needed for NGS data assembly and to access structural variation, i.e, contiguous genetic variation greater than 50 bp. The association of structural variation to various disease states is just now being appreciated and can be considered the next frontier for achieving comprehensive genetic understanding[25].

An additional newer effort in our group is the development of microfluidic CLIA-waivable (point-of-care) diagnostic devices. Our devices perform massively parallel ($10^6$) singleplex PCR or immnoPCR reactions on a device that accepts µL volume sample such as saliva or blood. Each reaction volume is capable of detecting a single molecule and thus performs digital assays. Detection limits for nucleic acids are single digit aM and for proteins single digit fM. Moreover, dynamic ranges up to $10^{10}$ have been demonstrated. This presentation will describe the early developments that lead into the commercialization of the first active control microfluidic device by Caliper Technologies where I was the sole scientific founder. Follow-on micro- and nanoscale devices developed by our group will be described and commercialization efforts associated with those technologies.